\documentstyle[12pt]{article}

\newcommand{\beg}{\begin{equation}}
\newcommand{\enq}{\end{equation}}
\newcommand{\sig}{\sigma}
\newcommand{\cst}{{\rm constant}}

\newcommand{\ga}{\gamma}

\newcommand{\Ga}{\Gamma}

\newcommand{\IpII}{{\frac{1-\ga}{2}}}
\newcommand{\pIII}{{\frac{1+\ga}{2}}}
\newcommand{\ro}{{\rho}}
\newcommand{\pri}{{\prime}}
\newcommand{\cS}{{{\cal S}}}
\begin{document}

\title{Gap probability of a one-dimensional gas model} 
\author{Y Chen$^{1,\dag}$ and S. M. Manning$^{2,\dag\dag}$\\
$^1$ Department of Mathematics, Imperial College\\
180 Queen's Gate,London SW7 2BZ UK\\
$^{2}$ Department of Theoretical Physics,Oxford University\\
1 Keble Road,OX1 3NP UK}
\date{\today}
\maketitle
\small{
\begin{abstract}
We investigate the gap formation probability of 
the effective one dimensional gas model recently proposed 
for the energy level statistics for disordered solids at the 
mobility edge. It is found that in order to get the correct form for
the gap probability of this model, the thermodynamic limit must be 
taken very carefully.
\end{abstract}}
\vskip 2cm
$^{\dag}$ y.chen@ic.ac.uk\\
$^{\dag\dag}$ s.manning1@physics.oxford.ac.uk
\vskip 1cm
{\bf To appear in the International Journal of Modern Physics B (1996).}
\vfill\eject
\noindent
{\bf Introduction}\\
The eigenvalue distribution of large random matrices has a long history
beginning
with the work of Wigner \cite{Wigner} on the energy level statistics of
heavy nuclei. 
This topic has found recent applications in the theory of disordered metals, 
quantum chaos
and string theory. The most studied and best understood of
 random matrix models
are the Gaussian ensembles with unitary, orthogonal and 
symplectic symmetries \cite{Mehta}.
Perhaps the most fundamental of all the statistical quantities of 
interest is
the probability that an appropriately scaled interval of the 
spectrum is free of eigenvalues. This is a measure of
the degree of correlation
between the eigenvalues. There is an ansatz of Wigner which gives 
a simple formula that
fits rather well the experimentally
observed nearest levels spacing distribution
\cite{Wigner} \footnote{Although
this ansatz deviates only minutely from results of 
numerical experiments on random matrices,
it is now known that the Wigner ansatz is not quite correct; 
the complete level 
spacing distribution functions are obtained from transcendental 
functions of various Painlev\'e types; the type 
depends the part of the spectrum where one performs
the scaling \cite{JMMS, Mehta1, TW}}. 
The energy levels of disordered metals
in the weak localization regime exhibit behavior which is rather 
well described by 
the correlation functions of the various Gaussian ensembles 
\cite{ASh,AS}. Furthermore the associated level spacing distribution 
function
compares favorably with the
Wigner Ansatz, but upon increasing the strength of disorder the level spacing
distribution evolves and deviates from the ``standard behavior.''
Perhaps 
the most striking result is that at the mobility edge the level spacing 
distribution
has a universal form when expressed in units of the average level 
spacings \cite{SSSLS}. A random matrix model has been proposed which 
accounts for this continuous deviation \cite{Chen1}. In this 
model the usual unitary Gaussian 
ensemble is $q-$ deformed in a similar way that to the
$q-$ deformation of the Laguerre ensembles \cite{Chen}. 
The results of this ensemble give qualitative features which are in
agreement with the
energy level statistics obtained from numerical work 
on disordered solids \cite{SSSLS}.
Using the two-level correlation function obtained in \cite{Chen1} 
it was established in \cite{Chen2} that the number variance and
the $\Delta_3$ statistics exhibit novel behavior. 
Very recently, using a combination
of scaling arguments and diagrammatical analysis, it has been 
suggested by the authors of \cite{i1} that the density of state 
correlation function 
in the bulk of spectrum
at the mobility edge behaves as $-1/|x-y|^{2-\ga}$ instead 
of $-1/|x-y|^2$ in the
regime of weak disorder. Here $x,\;y$ denotes the energy levels and
$\ga$ is related to the correlation length exponent $\nu$ and the
dimension ($d$) of the system: $\ga=1-\frac{1}{\nu d}.$\\ 
A gas
model---known as the effective plasma model in\cite{i2,i3}---with 
a power law pair
repulsion between the particles which gives 
correlation function: $-1/|x-y|^{2-\ga},$ was proposed.
Using a continuum 
approximation of Dyson \cite{Dyson}, the authors of \cite{i2,i3} 
conclude that there is 
a new universal law for the level spacing distribution function 
for the energy levels of a disordered metal at the mobility 
edge. This law states that the probability that the interval $(-s,s)$ of the
spectrum is free of levels behaves as  $E(-s,s)\sim {\rm e}^{-\cst\;s^{2-\ga}}
,$ where the $\cst$ appearing here is independent of $s.$\\
In the continuum limit, which is 
expected to be valid in the limit where $N$ (the total number of
particles) goes to infinity, 
the gas model may be described by a continuous
charge density $\sig,$ following the work of Dyson \cite{Dyson}.
In the mean field approximation this gas model has the
energy functional,
\beg
F[\sig]=\frac{1}{2}\int_{J}dx\int_{J}dy\frac{\sig(x)\sig(y)}{|x-y|^{\ga}}
+\int_{J}dx\sig(x)u(x)\;,\;\;\;\;0<\ga<1\enq subject to
$$\int_{J}dx\sig(x)=N,$$ where
$J$ is the support of the eigenvalues/fluid and
$u(x)$ is the confining potential that holds together the repelling
particles. Without loss of generality the coefficient of
$1/|x-y|^{\ga}$ in Eq. (1) is set to unity. This is denoted by $A_\ga$
in \cite{i2,i3}. Although the continuum approximation is not 
mathematically rigorous, previous experience suggests that it is 
rather robust \cite{Chen3}. The density that minimizes Eq.(1) satisfies,
\beg
\int_{J}dy\frac{\sig(y)}{|x-y|^{\ga}}=A-u(x),\;\;\;\;x\in J,\;\;\;\;
A={\rm Chemical\; potential}\enq supplemented by
$\int_{J}dx\sig(x)=N.$ At equilibrium the minimum free energy is
\beg
F[J]:=F_{\rm chem}+F_{\rm int},\enq where $F_{\rm chem}:=AN/2$ is
chemical potential contribution to the free energy and \\ $F_{\rm
int}:=\frac{1}{2}\int_{J}dx\sig(x) u(x),$ is the energy 
due to interaction between the charge distribution and the confining
potential. Eq.(3) can be obtained simply
by the use of the mean field equation that governs $\sig$ and the
normalization condition. 
Note that the separation into $F_{\rm chem}$ and $F_{\rm int}$
in Eq.(3) is not unique, different but equivalent expression can be 
found by using Eqns. (1) and (2) and the normalization condition.
To obtain the density we have to deal
with the inversion of the following integral equation:
\beg
\int_{J}\frac{\varphi(y)}{|x-y|^\ga}dy=\psi(x),\;\;\;x\in J.\enq
where $0<\ga<1,$ and $J$ is a subset of the real line. Assuming 
$\psi\in L^2[J],$ we seek $\varphi\in L^2[J].$ 
In the case where $J$ is a single
interval, say $J=(0,a),$ there is a unique inversion formula,
\beg
\varphi(x)=\frac{B_\ga}{x^{\frac{1-\ga}{2}}}\frac{d}{dx}
\int_{x}^{a}dt\frac{t^{1-\ga}}{(t-x)^{\frac{1-\ga}{2}}}
\;\frac{d}{dt}
\int_{0}^{t}ds\frac{\psi(s)s^{\frac{\ga-1}{2}}}
{(t-s)^{\frac{1-\ga}{2}}},\;x\in (0,a),
\enq
where $$B_\ga:=-\frac{\Ga(\ga)\cos\left(\frac{\pi\ga}{2}\right)}
{\pi\left[\Ga\left(\frac{1+\ga}{2}\right)\right]^2}.$$ 
Eq. (4) was investigated by
Carleman \cite{Carleman}, and later by 
Widom in the context of stable processes 
\cite{Widom}. This equation is also a special case of an integral 
equation with hypergeometric kernel \cite{Akhiezer}. 
However, in the many interval case,
$$J=\cup_{p=1}^n(a_p,b_p),\;\;\;a_1<b_1<\ldots<a_n<b_n,\;\;\;\;$$
no similar result has been obtained. This is an outstanding problem.\\ 
The quantity of interest, the
probability, $E[J],$ that an interval $J$ is free of
eigenvalues is 
$$
E[J]=
\frac{\left(\prod_{j=1}^N\int_{J^c}dx_j\right){\rm e}^{-W}}
{\left(\prod_{j=1}^{N}\int_{J\cup J^c}dx_j\right){\rm e}^{-W}},$$
where 
$$W=\sum_{j=1}^{N}u(x_j)+\sum_{1\leq j<k\leq N}
\frac{1}{|x_j-x_k|^{\ga}},$$
$J^c$ is the complement of $J$ and $J^c\cup J$ is the natural 
support of the particles.\\
From the previous equation we see that 
minus the logarithm of $E[J]$ is the change in
free energy; the free energy where all particles reside in $J^c$ minus
the free energy where all particles reside in $J^c\cup J,$ 
$$-\ln E[J]=\delta F:=F[J^c]-F[J\cup J^c].$$ 
The model considered in \cite{i2,i3}, corresponds to the particles
being distributed on the real line and and has the confining
potential $u(x)=x^2.$ 
The logarithm of the probability that the interval
$(-a,a)$ is free of particles is formally 
$$
-\ln E[(-a,a)]=\delta F:=F[(-b,-a)\cup(a,b)]-F[(-B,B)]$$
\beg
=\frac{1}{2}N
\left[A(a,N)-A(0,N)\right]+\frac{1}{2}\int_{(-b,-a)\cup(a,b)}
dx\sig(x,{\rm gap})u(x)-\frac{1}{2}\int_{-B}^{B}\sig(x)u(x),\enq
where
$b(>a>0)$ is band edge of the spectrum with a gap at $(-a,a)$
and $B$ is the band edge where the spectrum is gapless. We note that
with the normalization condition implemented, the respective edges $B$
and $b$ are functions of $N,$ and will diverge as
$N\rightarrow\infty.$ Furthermore, $b$ is also a function of $a.$
 The deliberate use of $B$ and $b$ to
distinguish the two situations is to indicate that although in the
thermodynamic limit, i.e.  $N\rightarrow\infty,$ $B$ and $b$ differs
by a very small amount their difference is nevertheless important and
can not be neglected, for in computing $\delta F$ we are subtracting
two very large numbers.  Concerning the
chemical potentials; $A(a,N)$ is the chemical potential where there 
is a gap and $A(0,N)$ is that without. 
The above remark also applies to the change in the chemical potentials.
The density,
$\sig(x,{\rm gap}),$ indicates that the integral equation must be
inverted in the interval $(-b,-a)\cup (a,b)$ for which an explicit
inversion formula is not known.\\ 
Although the change in the chemical
potential is expected to be small as $N\rightarrow\infty$ but due to
the multiplicative factor $N;$ this will give a significant contribution
to the change in free energy. The authors of \cite{i2,i3} assume 
that the change in the
chemical potential (denoted by $\mu$ in \cite{i2,i3}), 
$\delta\mu,$ is of order $s/{\cal E},$ (where
their $s$ is our $a$ and ${\cal E}$ is our $B$) and vanishes in the
thermodynamic limit, because ${\cal E}$ diverges as $N\rightarrow
\infty,$ giving negligible contribution to $\delta F.$ However,
further reflection suggests otherwise. From the symmetrical 
density given in \cite{i2,i3}, 
$\rho(x)\sim ({\cal E}^2-x^2)^{\frac{1+\ga}{2}},$ a simple calculation
using the normalization condition shows that ${\cal E}\sim
N^{\frac{1}{2+\ga}},$ thus $\delta F_{\rm
chem}\sim N\delta\mu\sim N^{\frac{1+\ga}{2+\ga}}s$ and 
cannot be neglected.
Indeed the scaling variable $\cS$ commonly used in random matrix
theory ( not the $s$ used in \cite{i2,i3}) 
contains the product of two
quantities one of which---$s$---tends to zero and the other a function
of $N$ tends to infinity. Therefore, the appropriate scaling
variable for the model studied in \cite{i2,i3} ought to be
$\cS=N^{\frac{1+\ga}{2+\ga}}s.$ In an example to be given below, where the
problem of multi-interval inversion of the integral equation is
circumvented, the meaning of the scaling variable $\cS$ will be
clarified in an exact computation within the mean field approximation.
In fact we will see that the change in chemical potential is of
order $a,$ and is {\it not} of order $a/B.$  
\\
\noindent
{\bf Gap formation probability of a semi-infinite model}\\
\noindent
As mentioned in the previous section in order to by-pass the problem
of multi-interval inversion, we consider the model with
$u(x)=x,\;x\geq 0.$ We are required to compute two free energies; one
with all the particles are confined in the interval 
$(a,b),{\rm where}\;b>a>0,$ and the
other with all the particles confined in $(0,B).$ Here $b$ is the
band edge for a density which has a gap in $(0,a),$ and is 
supported only
in $(a,b)$. The band edge $B$ is for a density that is supported over
$(0,B).$ Note that when the normalization condition is implemented $b$
will be a function of $N$ and $a$ while $B$ will be a function of $N$
only. The probability that the interval $(0,a)$ has no particle in the
thermodynamic limit is therefore
\beg
E[\cS]={\rm exp}\left[-\delta F\right],\enq where
\beg
\delta F:=F[(a,b)]-F[(0,B)].\enq
The scaling variable $\cS$ will become apparent later. To determine the
free energy, we first invert the integral equation,
\beg
\int_{a}^{b}dy\frac{\sig(y)}{|x-y|^{\ga}}=A-x,\;\;\;a<x<b,\enq
to find $\sig.$ With a shift in the variables
$x\rightarrow x+a,\;y\rightarrow y+a$ we can now use Eq. (5) to find,
for $a<x<b,$
\beg
\sig(x)=-\frac{B_\ga}{(x-a)^{\IpII}}\frac{\Ga^2\left(\pIII\right)}
{\Ga(\ga)}
\left[
\left(A-a-\frac{(b-a)(\ga+1)}{2\ga}\right)\frac{1}{(b-x)^{\IpII}}+
\frac{1}{\ga}(b-x)^{\pIII}\right].\enq
Note that since $0<\ga<1$ the density will have integrable divergences
at $a$ and $b.$ In order to have a configuration that has a lower free
energy we choose
\beg
A=(b-a)\left(\frac{1+\ga}{2\ga}\right)+a
\enq
 to eliminate the divergence at $b.$ This is in agreement with the
general theory of integral equations with weakly singular kernel. The
normalization condition, $\int_{a}^{b}dx\sig(x)=N$, produces,
\beg
N=\frac{\cos\left(\frac{\pi\ga}{2}\right)
\Ga^2\left(\pIII\right)}{2\pi \ga^2\Ga(\ga)}
(b-a)^{\ga+1}.\enq
From Eqns. (11) and (12), we find
\beg
b=\left(\frac{N}{C_\ga}\right)^{\frac{1}{\ga+1}}\;+a,\;\;\;
C_\ga:=\frac{\cos\left(\frac{\pi\ga}{2}\right)\Ga^2\left(\pIII\right)}{2\pi
\ga^2\Ga(\ga)}
\enq
and
\beg
A(N,a)=\left(\frac{\ga+1}{2\ga}\right)
\left(\frac{N}{C_\ga}\right)^{\frac{1}{\ga+1}}\;+a.\enq
Eq. (13) shows that $B=(N/C_\ga)^{\frac{1}{\ga+1}}.$
The first term in the r.h.s. of Eq. (14) is identified as $A(N,0),$ 
the gapless chemical potential. The change in the chemical potential is 
therefore $a$ and {\it not} $a/B$ as claimed in \cite{i2,i3}.  
With $\sig$ given by Eq. (10) and $A(N,a)$ given by Eq. (14), we 
compute the free energy in the presence of the gap,
using elementary integration formulae,
\beg
F_{\rm int}=\frac{Na}{2}+
\frac{\Ga^2\left(\frac{\ga+3}{2}\right)
\cos\left(\frac{\pi\ga}{2}\right)}
{2\pi \ga\Ga(\ga+3)}
\left(\frac{N}{C_\ga}\right)^{\frac{\ga+2}{\ga+1}},\enq
\beg
F_{\rm chem}=\frac{\ga+1}{4\ga}N
\left(\frac{N}{C_\ga}\right)^{\frac{1}{\ga+1}}+\frac{Na}{2}.\enq
Therefore the free energy for which all particles are confined in
$(a,b)$ is
\beg
F(a,b)=Na+f_\ga(N),\enq where
\beg
f_\ga(N):=\frac{\ga+1}{4\ga}N
\left(\frac{N}{C_\ga}\right)^{\frac{1}{\ga+1}}+
\frac{\Ga^2\left(\frac{\ga+3}{2}\right)
\cos\left(\frac{\pi\ga}{2}\right)}
{2\pi \ga\Ga(\ga+3)}
\left(\frac{N}{C_\ga}\right)^{\frac{\ga+2}{\ga+1}}.\enq
We identify $f_\ga(N)$ with $F(0,B)$, the free energy for which all
particles are confined in $(0,B)$ and the change in free energy is
exactly,
\beg
\delta F(0,a)=F(a,b)-F(0,B)=Na.\enq
The limit which produces the scaling variable is defined as
$a\rightarrow 0$ and $N\rightarrow\infty$ such that $\cS:=Na$ is
finite. Thus the probability that the interval $(0,a)$ is free of
particles is
\beg
E[(0,a)]={\rm e}^{-\cS}.\enq 
To further illustrate the meaning of ${\cal S},$ we
consider another model with $u(x):=x^2,\;x\geq 0.$ 
Using Eq.(5), and after some computation, we find for
$a<x<b,$
\beg
\sig(x)=-\frac{\cos\left(\frac{\pi\ga}{2}\right)}{4\pi \ga(\ga+1)}
\frac{1}{\left[(x-a)(x-b)\right]^{\IpII}}
\left[C+\left[4a-12b-4(a+b)\ga\right](b-x)+8(b-x)^2\right],\enq
where
\beg
C:=-a^2-2ab+3b^2-4A\ga(\ga+1)+4b^2\ga+(a+b)^2\ga^2.\enq 
Just as in the previous example we choose
\beg
A=\frac{(a+b)^2\ga+3b^2-a^2-2ab}{4\ga}
\enq
so that $C=0,$ to eliminate the integrable singularity of 
$\sig$ at $b.$ Thus
\beg
\sig(x)=\frac{\cos\left(\frac{\pi\ga}{2}\right)}{\pi \ga(\ga+1)}
\frac{(b-x)^{\pIII}}
{(x-a)^{\IpII}}\left[(\ga+3)b-(1-\ga)a-2(b-x)\right],\;\;a<x<b.\enq
We note that the factor $[\ldots]$ in Eq.(24) for $a<x<b$ is greater
than $(a+b)(\ga+1)$ and is therefore positive. Application
of the normalization condition leads to a transcendental equation for
$b,$
\beg
N=\frac{D_\ga\Ga^2\left(\pIII\right)}{2\Ga(\ga+1)}(b-a)^{\ga+2}
\left[1+\frac{\ga(b+a)}{b-a}+\frac{\ga+1}{\ga+2}\right],\;\;
D_\ga:=\frac{\cos\left(\frac{\pi\ga}{2}\right)}{\pi \ga(\ga+1)}.
\enq 
Although the
explicit solution of the transcendental equation governing $b$ is not
known we may nevertheless determine $b$ as a power series in $a,$
which in turn can be used to calculate $F_{\rm int}$ and $F_{\rm
chem},$ also in a power series in $a.$ After some straightforward but
lengthy calculations not reproduced here, we find
\beg
\delta F\sim aN^{\frac{3+\ga}{2+\ga}}\left[1+{\rm O}
\left(\frac{a}{N^{\frac{1}{2+\ga}}}\right)\right].\enq
In this case the appropriate scaling variable is
\beg
\cS:=aN^{\frac{3+\ga}{2+\ga}},\enq
and
\beg E[(a,b)]\sim {\rm e}^{-\cS}.\enq
We see that although the scaling variable $\cS$ is the same in both
cases, the thermodynamic limit is arrived in a {\it distinct} manner
for distinct potentials. Following the procedure described in
\cite{i3} we will find $-\ln E[s]\sim s$ for our potentials. 
To see how
this comes about, we compute the change in the free energy according to
the procedure given in \cite{i3} for the semi-infinite models
studied above. In order
to determine the probability that there are no particles in $(0,s)$,
we require the change in free energy, which according to \cite{i3},
is
\beg
F_s-F_0=-\frac{1}{2}\int_{s}^{\infty}dx\int_{s}^{\infty}dx^{\prime}
\frac{\delta\ro_s(x)\delta\ro_s(x^{\prime})}{|x-x^{\prime}|^{\ga}} 
+\frac{1}{2}\int_{0}^{s}dx\int_{0}^{s}dx^{\pri}
\frac{\ro_0(x)\ro_0(x^{\pri})}{|x-x^{\pri}|^{\ga}},\enq 
where, $\delta\ro_s(x)$, the change in the density due to the 
presence of the gap is related to the density in the absence of the gap,
$\ro_0(x),$ via the following integral equation:
\beg
\int_{s}^{\infty}dx^{\pri}\frac{\delta\ro_s(x^{\prime})}
{|x-x^{\pri}|^{\ga}}=\int_{0}^{s}dx^{\pri}\frac{\ro_0(x^{\pri})}
{|x-x^{\pri}|^{\ga}}.\enq 
Eq. (30) which governs $\delta\ro_s$ and which provides a determination
of $F_s-F_0$ is equivalent to the statement that the change in the
chemical potential vanishes in the thermodynamic limit \cite{i3}.
However, as noted from the examples given above that 
the change of chemical potential is of order $a$. 
We now deduce the $s$ dependence of the gap formation probability 
from Eqns. (29) and (30). With the change of variables used in \cite{i3}:
$z=s/x,\;\;t=s/x^{\pri},$ and $u_g(z)=|z|^{-2+\ga}\delta\ro_s(s/z),$
Eq. (30) becomes,
\beg
\int_{0}^{1}dt\frac{u_g(t)}{|z-t|^{\ga}}
=\frac{1}{s^{\frac{1-\ga}{2}}}
\int_{1}^{\infty}\frac{dt}{t^{\frac{3-\ga}{2}}|z-t|^{\ga}}.\enq
To obtain this we have used the fact the integral in the r.h.s. of 
Eq. (30) is 
dominated by $\ro_0(x)$ near the origin: $\ro_0(x)\sim
1/x^{\frac{1-\ga}{2}}$. Note that the $t$ integration in the r.h.s. of 
Eq.(31) converges. From the Eq. (31) we see that the
combination $u_g(t)s^{\frac{1-\ga}{2}}$ is {\it independent} of $s.$
Using this, and the variables $z$, $t$ and $u_g(t),$ we find from a
simple dimensional argument, $$F_s-F_0=\cst\;\; s,$$ where
the constant is {\it independent} of $s.$ 
This suggests that the change in the free energy 
is independent of the confining potential $u(x)$ and universal. 
The statement of universality requires some qualification. If the
confining potential $u(x)$ is far weaker then those considered
above, the singularity of density near the origin, 
(which is determined uniquely by the confining potential via the
integral equation,) 
will be modified and we may expect the final form of $\ln E[s]$ to
be modified accordingly.\\
It is interesting to note that for sufficiently strong confining 
potentials the singularity of the density is universal, i.e., $\rho_0(x)\sim
1/x^{\frac{1-\ga}{2}}.$ This  phenomena is also observed in the
standard random matrix ensembles delineating the behaviour of the
density for strong and weak confining potentials (no longer 
universal) \cite{Chen,Chen1,Chen2,Chen3}. Thus in the
situations where the singularity structure of the density
is universal a universal form for $-\ln E[s]$ is found.\\
It appeared
by adopting the two distinct approaches; 
one of which computes the change in the chemical potential explicitly 
by keeping $N$ finite but large while the other taking the limit
$N\to\infty$ right off the start, we are led the same law for the
gap probability, but with distinct scaling variable ${\cal S}=N^{\nu}a$
as opposed to $s.$ It may be of interest to understand better the 
thermodynamic limits and clarify the meaning of $s$ and ${\cal S}.$
\\
We are at present developing tools in an attempt to obtain a
multi-interval inversion formula.
\vskip .5cm
{\bf Acknowlegement}\\
The authors should like to thank Mourad Ismail and Alex Hewson 
for helpful discussion.
\vfill\eject

\end{document}